УДК 519.876.5

В.М. Дмитриев, Т.В. Ганджа, В.В. Ганджа, С.А. Панов

# Структура и функции автоматизированной системы управления проектами для центров научно-технического творчества студентов

Рассматриваются возможности автоматизации студенческого проектирования (выполняемого в рамках центров научно-технического творчества студентов) с помощью использования автоматизированной системы управления проектами. Описаны назначение, состав и формализм автоматизированного рабочего места студента-проектировщика, а также приведена его структурно-функциональная схема.
**Ключевые слова:** проект, система, управление, система управления, автоматизация, проектирование.

В 2012 г. в ТУСУРе был создан центр научно-технического творчества студентов (ЦНТТС), основная цель которого заключается в реализации творческих способностей студентов, обучающихся по техническим специальностям. Особое внимание при создании ЦНТТС было уделено такому направлению деятельности студентов, как студенческое проектирование. Студенческое проектирование – это образовательная и первоначальная научно-исследовательская деятельность, которая представляет собой разработку управляемых технических устройств и систем. С целью автоматизации, управления и сопровождения процесса студенческого проектирования в ЦНТТС разрабатывается автоматизированная система управления проектами (АСУПР). Разработка АСУПР позволит:
– автоматизировать процесс студенческого проектирования;
– существенно упростить процедуру координации руководителей и исполнителей проектов;
– отслеживать ход выполнения проектов;
– предоставлять доступ к уже завершенным проектам с целью их дальнейших доработок.

В качестве объектов проектирования в ЦНТТС при использовании АСУПР выступают кибермодели, которые являются масштабированной копией мультифизических сложных технических устройств или систем (СТУС): управляемых игровых моделей, бытовых приборов, промышленного и военного оборудования. Предназначение этих кибермоделей – практическое изучение с различной степенью детализации процессов, протекающих в реальных устройствах. Проект в АСУПР реализуется с помощью программно-аппаратного комплекса, имеющего название «Автоматизированное рабочее место студента-проектировщика» (АРМСП).

Чтобы сформулировать принципы функционирования и задачи, решаемые АРМСП, сначала необходимо рассмотреть обобщенную структуру СТУС.

**Формализованное представление объекта исследования и проектирования при работе на АРМСП.** Объект исследования обозначен как мультифизические (т.е. с компонентами различной физической природы) СТУС. В его состав могут входить элементы электроники, механики и электромеханики, а также блоки измерения и управления. К СТУС, например, относятся электромеханические и мехатронные системы, в которых обеспечение движения является целью управления. Это также робототехнические системы, строительные и дорожные машины и механизмы, а также электропривод различной аппаратуры из области приборостроения. К классу СТУС как объекта исследования также относятся и электрические системы – энергетические системы генерации и преобразования электрической энергии.

С позиций общего подхода к моделированию и анализу СТУС может быть названа совокупность объектов «источник – преобразователь – нагрузка – измерение – управление».

С учетом вышесказанного обобщенная функциональная схема СТУС, задачей которой является обеспечение движения, представлена на рис. 1. В ее состав входят один или несколько двигателей с совокупностью передаточных устройств и необходимым рабочим оборудованием, силовой преобразователь, микропроцессорная система управления и информационно-измерительная система. По входу СТУС взаимодействует с первичным источником энергии и человеком оператором, а по выходу – с нагрузкой.

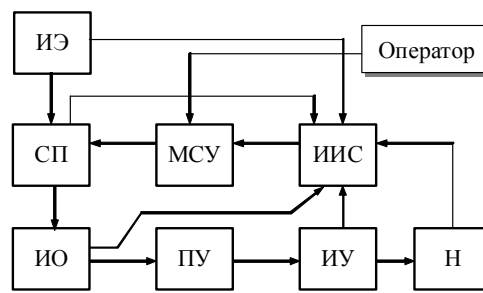

Рис. 1. Обобщенная схема СТУС: ИЭ – источник энергии; ИО – исполнительный орган; Н – нагрузка; ПУ – передаточное устройство; СП – силовой преобразователь; МСУ – микропроцессорная система управления; ИУ – исполнительное устройство; ИИС – информационно-измерительная система

Создание АРМСП для таких объектов, как СТУС (см. рис. 1), представляется на данный момент чрезвычайно трудной задачей, подойти к решению которой целесообразней постепенно, рассмотрев сначала более узкий класс объектов, например электротехнических систем (ЭТС) или систем автоматического управления (САУ), для которых имеются уже отлаженные и апробированные системы расчета, моделирования и автоматизации эксперимента.

Поэтому на базе обобщенной схемы могут быть синтезированы и другие более простые варианты СТУС и их подсистем, например для статических систем, в функции которых не входит обеспечение движения.

На основе рассмотренной структуры СТУС можно выделить основные функции АРМСП.

**Основные функции автоматизированного рабочего места студента-проектировщика.** АРМСП выполняет следующие функции:

− Формирование технического задания на разрабатываемый объект, выявление основного назначения разрабатываемого объекта, его требуемых технических и тактико-технических характеристик, показателей качества и технико-экономических требований.

− Формирование структуры проектируемого СТУС, выявление его компонентного состава, а также значений параметров компонентов, удовлетворяющих предписанным в техническом задании характеристикам, показателям качества и технико-экономическим требованиям.

− Формирование оптимальной конструкции проектируемого объекта или устройства, взаимного расположения его функциональных блоков в ограниченном геометрическом пространстве, разработка печатных плат электронных блоков и мест крепления радиокомпонентов.

− Разработка принципов и последовательности действий, направленных на создание возможностей тиражирования проектируемого объекта. Формирование технологических карт работы и программ для станков с числовым программным управлением (ЧПУ).

− Разработка экспериментального образца, исследование и проверка его свойств и характеристик (на соответствие техническому заданию).

− Принятие решений по управлению кибермоделями.

**Структурно-функциональная схема АРМСП.** С целью реализации вышеперечисленных функций была предложена оптимальная структурно-функциональная схема АРМСП (рис. 2).

Приведенная выше структура АРМСП включает в себя следующие основные элементы:

1. **Интерактивное техническое задание на проект** – это документ, предписывающий проектанту исходные требования, условия, характеристики и ограничения на создаваемый продукт. Должно быть выполнено в форме интерактивного шаблона, допускающего занесение и редактирование исходных данных, заданных в форме: текста, цифровых соотношений и формул, расчетных выражений, рисунков и схем, таблично-графических данных.

2. **Система поиска прототипов** – программа, служащая для поиска аналогов объектов проектирования, обладающих близкими техническими характеристиками, заданными в табличной форме, и подлежащих сравнению с таблично-заданным техническим заданием на проект.

3. **Банк проектных решений** – специализированная база данных, обеспечивающая систематизацию сведений об объектах разработки (схемы, параметры, режимы) и автоматизированный выбор аналога.

4. **Банк справочных материалов** – специализированная база данных, предназначенная для хранения и поиска параметров элементов систем: типовых узлов и деталей аппаратуры (диодов, транзисторов и др.). Банк справочных материалов непосредственно связан с интернет-ресурсом.

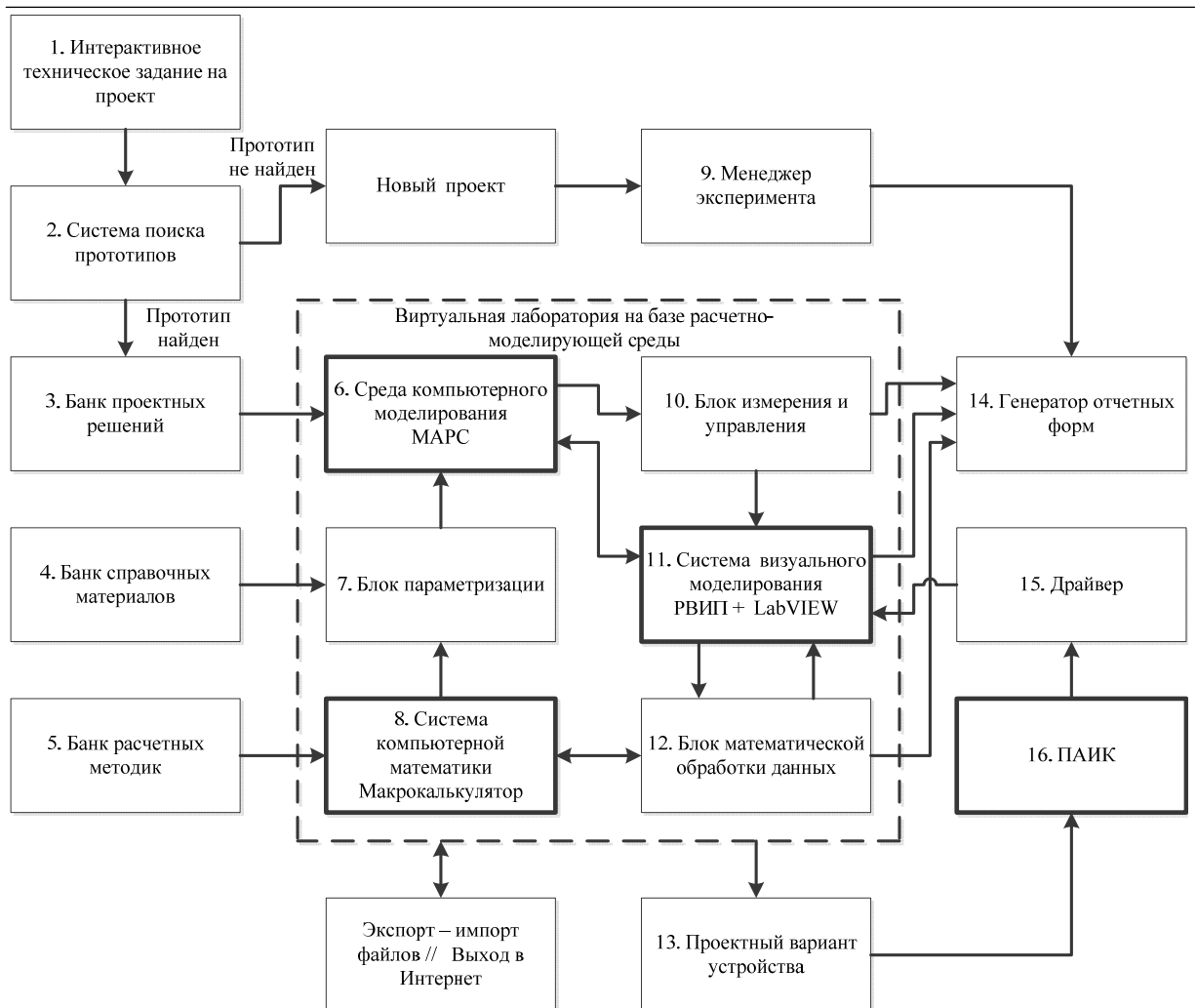

Рис. 2. Структурно-функциональная схема АРМСП

5. **Банк расчетных методик** – специализированная база данных, обеспечивающая систематизацию инженерных методик, использующихся в процессе проектирования, сопровождение банка и выполнение необходимых расчетов в системе компьютерной математики «Макрокалькулятор».

6. **Среда моделирования МАРС** [1] – комплекс алгоритмов и программ, используемый для компьютерного моделирования проектного варианта устройства.

7. **Блок параметризации** – программное средство, позволяющее на основе исходных данных произвести вычисления значений необходимых параметров (компонентов объектов разработки) и транслировать их в компьютерную модель.

8. **Система компьютерной математики «Макрокалькулятор»** [2] – программное средство, используемое для вычисления значений параметров компьютерной модели проектного варианта устройства, а также для реализации сложных алгоритмов обработки выходных данных.

9. **Менеджер эксперимента** – программное средство, обеспечивающее автоматизированное управление ходом выполнения эксперимента.

10. **Блок измерения и управления** – программное средство, обеспечивающее автоматизацию измерений и управления данными, получаемыми в результате компьютерного моделирования, и осуществляющее передачу этих данных в другие блоки.

11. **Система визуального моделирования РВИП+LabVIEW** [3] – служит в АРМСП для разработки виртуальных приборов; обработки данных; управления вычислительным экспериментом.

12. **Блок математической обработки данных** – программное средство, обеспечивающее автоматизированный расчет данных, получаемых из других блоков.

13. **Проектный вариант устройства** – спроектированный вариант устройства, подлежащий проверке на физическом макете.

14. **Генератор отчетных форм** [4] – служит для автоматизированного создания отчетов, содержащих результаты моделирования, анализа и синтеза проектируемых устройств.

15. **Драйвер** – программа, отвечающая за взаимодействие программных и аппаратных средств.

16. **ПАИК** – программно-аппаратный измерительный комплекс «Лабораторное автоматизированное рабочее место» (ЛАРМ или ЭЛВИС), к которому подключается физический макет.

17. Особое внимание при разработке структуры АРМСП было уделено применению среды моделирования МАРС и системы компьютерной математики «Макрокалькулятор», основанных на универсальном методе компьютерного моделирования СТУС – методе компонентных цепей, что позволило полностью автоматизировать процесс проектирования кибермоделей на всех его этапах.

**Заключение.** При разработке АСУПР и АРМСП была достигнута основная цель – автоматизация всех шагов проектирования кибермоделей, создаваемых школьниками и студентами первых курсов в рамках деятельности центров научно-технического творчества студентов (ЦНТТС).

**Дмитриев Вячеслав Михайлович**
Д-р техн. наук, профессор, декан факультета моделирования систем ТУСУРа
Тел.: (382-2) 41-39-15
Эл. почта: decan@toe.tusur.ru

**Ганджа Тарас Викторович**
Канд. техн. наук, доцент каф. системного анализа ТУСУРа
Тел.: (382-2) 41-39-15
Эл. почта: gandgatv@gmail.com

**Ганджа Василий Викторович**
Аспирант каф. моделирования и основ теории цепей ТУСУРа
Тел.: (382-2) 41-39-15
Эл. почта: vasivik@gmail.com

**Панов Сергей Аркадьевич**
Аспирант каф. системного анализа ТУСУРа
Тел.: (382-2) 41-39-15
Эл. почта: spytech3000@gmail.com

Dmitriev V.M., Gandzha T.V., Gandzha V.V., Panov S.A.
**The structure and functions of an automated project management system for the centers of scientific and technical creativity of students**

This article discusses the possibility of automating of the student's projecting through the use of automated project management system. There are described the purpose, structure and formalism of automated workplace of student-designer (AWSD), and shown its structural-functional diagram.
**Keywords:** project, system, control, control system, automation, design.